# MOS-2: A Two-Dimension Space for Positioning MAS Organizational Models

Hosny Abbas[1*], Samir Shaheen[2]

[1] Department of Electrical Engineering, Assiut University, Assiut, Egypt,
Email: hosnyabbas@aun.edu.eg

[2] Department of Computer Engineering, Cairo University, Giza, Egypt,
Email: sshaheen@eng.cu.edu.eg

**Abstract-** The increased complexity and dynamism of present and future Multi-Agent Systems (MAS) enforce the need for considering both of their static (design-time) and the dynamic (run-time) aspects. A type of balance between the two aspects can definitely give better results related to system stability and adaptivity. MAS organization is the research area that is concerned with these issues and it is currently a very active and interesting research area. Designing a MAS with an initial organization and giving it the ability to dynamically reorganize to adapt the dynamic changes of its unpredictable and uncertain environment, is the feasible way to survive and to run effectively. Normally, MAS organization is tackled by what is called, MAS organizational models, which are concerned with the description (formally or informally) of the structural and dynamical aspects of agent organizations. This paper proposes a two-dimension space, called MOS-2, for positioning and assessing MAS organizational models based on two dimensions: their adopted engineering viewpoint (agent-centered or organization-centered) as the vertical dimension and the agents awareness/unawareness of the existence of the organizational level as the horizontal dimension. The MOS-2 space is applied for positioning a number of familiar organizational models. Its future trends and possible improvements are highlighted. They include the following, (1) adding Time as a dimension, (2) increasing the considered dimensions, (3) providing a quantitative approach for positioning MAS organizational models.

## 1. INTRODUCTION

In contrast to initial MAS research, which concerned individual agents' aspects such as agents' architectures, agents' mental capabilities, behaviors, etc, the current research trend of MAS is actively interested in the adaptivity, environment, openness and the dynamics of these systems. In open environments, agents must be able to adapt towards the most appropriate organizations according to the state of the environment, which changes in an unpredictable manner. MAS organization [1] is currently considered as an emergent area of MAS research that relies on the notion of openness and heterogeneity of MAS and imposes new demands on traditional MAS models [2]. Considering MAS with no real structure isn't suitable for handling current software systems complexity, and higher order abstractions should be used and some way of structuring the society is typically needed to reduce system complexity, to increase system efficiency, and to more accurately model the problem being tackled [3].

Horling et al. [4] stated that our real world getting more complex and highly distributed and that should be reflected in new software engineering paradigms such as MAS. Therefore, the adoption of higher order abstract concepts like organizations, societies, communities, and groups of agents can reduce complexity, increase efficiency, and improve system scalability. Shehory [5] defined MAS organization as the way in which multiple agents are organized to form a multi-agent system. The relationships and interactions among the agents and specific roles of agents within the organizations are the focus of MAS organization. Dignum [26] pointed out that MAS organization can be understood from two perspectives, (1) organization as a process, and (2) organization as an entity. The first perspective considers agents organization as the process of organizing a set of individual agents, thus in this sense it is used to refer to constraints (structures, norms and patterns) found in a social context that shape the actions and interactions of agents. On the other hand, the second perspective considers agents organization as an entity in itself, with its own requirements and objectives and is represented by (but not identical to) a group of agents. In fact, MAS organization demands the integration of both perspectives and relies for a great extent on the notion of openness and heterogeneity of MAS.

There are two familiar viewpoints of MAS engineering, the first one is the agent-centered MAS (ACMAS) in which the focus is given to individual agents. With this viewpoint, the designer concerns the local behaviors of agents and also their interactions without concerning the global structure of the system. The global required function of the system is supposed to emerge as a result of the lower level individual agents interactions. The agent-centered approach takes the agents as the "engine" for the system organization, and agent organizations implicitly exist as observable emergent phenomena, which states a unified bottom-up and objective global view of the pattern of cooperation between agents [6]. Ant colony [7] is a natural example of the ACMAS viewpoint, where there is no organizational behavior and constraints are explicitly and directly defined inside the ants. The main idea is that the organization is the result of the collective emergent behavior due to how agents act their individual behaviors and interact in a common shared and dynamic environment. In ACMAS, the MAS organization is actually a process not an entity; there is a consensus to call this process as *self-organization* [28][29][30].

The second viewpoint of MAS engineering is what is called organization-centered MAS (OCMAS) in which the structure of the system is given a bigger attention through the explicit abstraction of agent organizations. With that approach, the designer designs the entire organization and coordination patterns on the one hand, and the agents' local behaviors on the other hand. It is considered as a top-down approach because the organization abstraction imposes some rules or norms used by agents to coordinate their local behaviors and interactions with other agents. In OCMAS, the MAS organization is actually an explicit entity not a process and to distinguish it from the ACMAS approach, the change in system organization is often called *dynamic reorganization* [6][24], which is a more general name than self-organization.

When a researcher proposes an approach to dynamically reorganize a multi-agent system to adapt environments' changes, he actually proposes what the MAS community agreed to call it as an *organizational model* [19]. MAS organizational models will play a critical role in the development of future larger and more complex MAS. The main concern of organizational models is to describe the structural and dynamical aspects of organizations [8]. They have proven to be a useful tool for the analysis and design of multi-agent systems. Furthermore, they provide a framework to manage and engineer agent organizations, dynamic reorganization, self-organization, emergence, and autonomy within MAS.

Picard et al. [6] added the agents' awareness /unawareness of the existence of the organization structure as a dimension of the organization modification process and he identified four cases, (1) the agents don't represent the organization, although the observer can see an emergent organization. In some sense, they are unaware that they are part of an organization, (2) each agent has an internal and local representation of cooperation patterns which it follows when deciding what to do, this local representation is obtained either by perception, communication or explicit reasoning, (3) the organization exists as a specified and formalized schema, made by a designer but agents don't know anything about it and even do not reason about it. They simply comply with it as if the organizational constraints were hard-coded inside them, (4) agents have an explicit representation of the organization which has been defined, the agents are able to reason about it and to use it in order to initiate cooperation with other agents in the system.

In this paper, we propose a two-dimension space inspired from a previous work of Picard et al. [6], for positioning and comparing MAS organizational models. The proposed space is similar to a two-dimension Cartesian coordinate system, where the ACMAS/OCMAS are represented by the vertical axis and the agents' awareness/unawareness of the existence of the organizational level are represented by the horizontal axis as shown in Figure 1. A MAS organizational model is represented as a point (small circle) or an area (oval) as demonstrated in the figure and as will be explained later. A number of familiar organizational models are positioned and compared using the proposed MAS organization space; called MOS-2, where MOS stands for **MAS O**rganization **S**pace and the number 2 indicates that the space is a two dimension space.

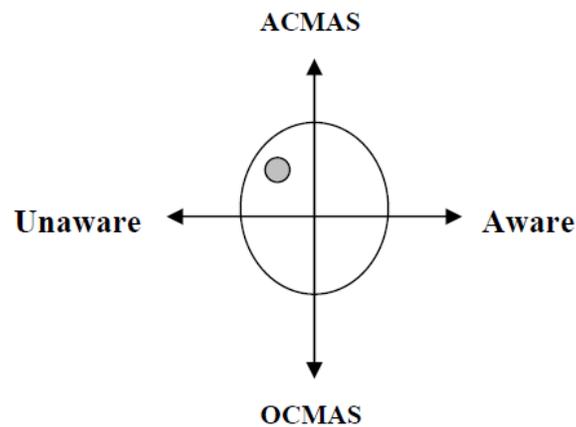

Figure 1: The MOS-2 MAS Organization Space

The central circle shown in the MOS-2 space can be seen as the unity circle (its radius is one). By this way it will be possible to precisely position an organizational model by determining the extent (i.e., percentage) to which it is an ACMAS or OCMAS approach and also the extent to which the individual agents are aware or unaware about the organizational aspects. In other words, the MOS-2 space can be used quantitatively not just qualitatively; this point is left as a future work but will be highlighted later in this paper. Furthermore, in the proposed space, the number of dimensions is two (Aware/Unaware, ACMAS/OCMAS), but it is also possible to increase the considered dimensions by adding new comparison aspects to address a fine-grained classes of MAS organizational models. In case the number of the considered comparison aspects (dimensions) is increased to be N, then the resultant space will be an N dimension space and can be called, MOS-N. Actually, this paper proposes only the two dimension case (MOS-2); other versions of higher dimensions are left as a future work.

The remaining of this paper is organized as follows: Section 2 explores the related work. Section 3 provides a background related to the two adopted comparison aspects. Section 4 presents the proposed MOS-2 space for positioning MAS organizational models. Section 5 demonstrates the applicability of MOS-2 on a number of familiar MAS organizational models and approaches. Section 6 discuss the results and provides a complete view based on these results. Section 7 highlights the possible future trends. Finally, Section 8 concludes this paper.

## 2. RELATED WORK

MAS organization was and still a very active and interesting research point in MAS. Concepts like organization, dynamic reorganization, self-organization, and emergence have attracted great attention in the last few years. The reason is related to the increasing complexity and highly distribution of modern real-life applications. This paper is not aimed to explain in details these concepts but interested readers can inspect their related references, for example [22][23][24][25][26][7][27]. The goal of this paper is to propose a visual semi-formal method for positioning and comparing a broad band of MAS organizational models. Organizational models are the tools to design methods, techniques, and approaches for managing the static and dynamic organization of MAS. Many researchers provided valuable narrative surveys and reviews contain informal analysis and evaluation of MAS organizational models. For instance, Picard et al. [6] aimed to study and propose a comprehensive view of how one could make multi-agent organizations adapted to dynamics, openness and large-scale environment. The authors proposed an analysis grid of different MAS organization approaches. The proposed grid has two dimensions: the vertical dimension identifies if the considered organization approach adopts the ACMAS, OCMAS, viewpoints. The second dimension is concerned with the awareness/unawareness of the individual agents of the organizational aspects. The authors claimed that the two dimensions of their grid are continuous, and it is completely possible to identify approaches that are at the boundary of two categories. Our propose work is inspired from Picard et al. work but in a more formal, visual and interesting way.

Alberola [18] provided, as a part of her PhD thesis, an analysis of how current reorganization approaches in MAS provide support to agent designers in order to develop adaptive agent societies. She described in detail some of the most relevant existing approaches, in order to show the advantages and limitations of each one. Alberola suggested that reorganization in agent societies can be represented as a loop process composed by different phases: Monitor, Design, Selection, and Evaluation phases. She also studied a number of familiar MAS organizational models by identifying the techniques adopted in each phase for each organizational model. Alberola study is valuable and it benefits us a lot but it is informal and contains intensive information and that makes it difficult to be captured by students and beginners.

V. Dignum [19] edited a handbook of research in MAS aimed to provide an overview of current work in agent organizations, from several perspectives, and focus on different aspects of the organizational spectrum. The handbook explored a number of familiar MAS organizational models and what makes it interesting is that the authors of the selected models wrote themselves the chapter that tackled their model. From the other hand, the handbook did not provide a general comparison or any type of positioning of the considered models.

Jensen et al. [20] investigated the agent-centered and organization-centered approaches to designing and implementing multi-agent systems. The authors have developed and evaluated two teams of agents for a variant of the well-known Bomberman computer game. One team is based on the basic Jason system, which is an implementation in Java of an extension of the logic-based agent-oriented programming language AgentSpeak. The other team is based on the organizational model Moise+, which is combined with Jason in the middleware called J-Moise+. They concluded that the agent-criented approach has a number of advantages when it comes to game-like scenarios with just a few different character types.

Horling and Lesser [21] also stated that organizational design employed by an agent system can have a significant, quantitative effect on its performance characteristics, and they surveyed the major organizational paradigms used in multi-agent systems. These include hierarchies, holarchies, coalitions, teams, congregations, societies, federations, markets, and matrix organizations. Also, they provided a description of each paradigm, and discuss its advantages and disadvantages, further, they provided examples of how each organization paradigm may be instantiated and maintained. But their work was not targeted to organizational models, which concerns both of static and dynamic aspects; they just concerned how to structure MAS with different paradigms.

In nutshell, the related work was valuable for us in designing the MOS-2 space for positioning MAS organizational models and approaches in a visual, semi-formal, easy to understand way. Providing a 2-dimension space for identifying an organizational model features and limitations in the scope of two or more dimensions is a good idea. It provides an effective tool to compare visually an organizational model with other models. This way enables designers and beginners to quickly capture a certain model in their minds and allows them to remember easily the considered model and its features or limitations relative to other models.

## 3. BACKGROUND

This paper is not concerned with the promotion of one MAS engineering viewpoint (ACMAS or OCMAS) relative to the other one, but the main concern is to position MAS organizational models and show the extent to which a certain model benefits from the adoption of each of these viewpoints. We have to emphasize here that both of the MAS engineering viewpoints (ACMAS/OCMAS) are generally not mutually exclusive and have led to different approaches in the domain [6]. In other words, it is possible to mix both viewpoints in one organizational model to take benefit of their pros and avoid their cons. Table 2 provides a comparison between the two viewpoints by presenting the characteristics and the shortcomings of each one. Also, Figure 2 provides the advantages and disadvantages of the adoption of each viewpoint relative to the individual agents' awareness or unawareness of the higher level organizational aspects.

Figure 2 represents the basis of the proposed two-dimension space as it orthogonally aligns the two comparison aspects into a vertical (ACMAS/OCMAS) and a horizontal (Awareness/Unawareness) axes. As demonstrated in the figure, four quadrants are resulted: ACMAS-Awareness, ACMAS-Unawareness, OCMAS-Unawareness, and OCMAS-Awareness. Therefore, a MAS organizational model that is positioned in one of these quadrants will simultaneously benefit and suffer from the advantages and disadvantages of this quadrant respectively. Note also that, an organizational model can be positioned into more than one quadrant.

Table 1: Comparison between the ACMAS and OCMAS viewpoints

|  | **Characteristics** | **Shortcomings** |
|---|---|---|
| **ACMAS** | <ul><li>Organization is a process (self-organization)</li><li>Informal/Bottom-up/Emergent/Endogenous</li><li>The focus is given to individual agents</li><li>Agents are the "engine" for the system organization</li><li>An agent may communicate with any other agent</li><li>An agent is responsible to define its relations with other agents</li><li>An agent is responsible to constrain its accessibility from other agents</li><li>Agents are autonomous and no constraint is placed on the way they interact</li><li>An agent provides a set of services available to other agents</li></ul> | <ul><li>Unpredictability and Uncertainty: lead to unreliability</li><li>Lack of Modularity: all agents are accessible from everywhere</li><li>Undesirable Emergent Behavior: can impact system performance</li><li>Dual Responsibility: agents have to manage simultaneously both the functional an the organizational aspects</li></ul> |
| **OCMAS** | <ul><li>Organization is often an explicit entity</li><li>Support dynamic reorganization</li><li>Formal/ To-down/ Pre-exist organization</li><li>Reduce system complexity</li><li>Increase system efficiency</li><li>Improve system scalability</li><li>Provide Effective coordination</li><li>Limit the scope of interactions</li><li>Tuning of the agents autonomy</li><li>Structuring of agents interactions</li><li>Separation of concerns</li><li>Modularity/Reliability</li></ul> | <ul><li>Computational / Communication overhead</li><li>Reduce overall flexibility or reactivity</li><li>Add additional layer of complexity</li></ul> |

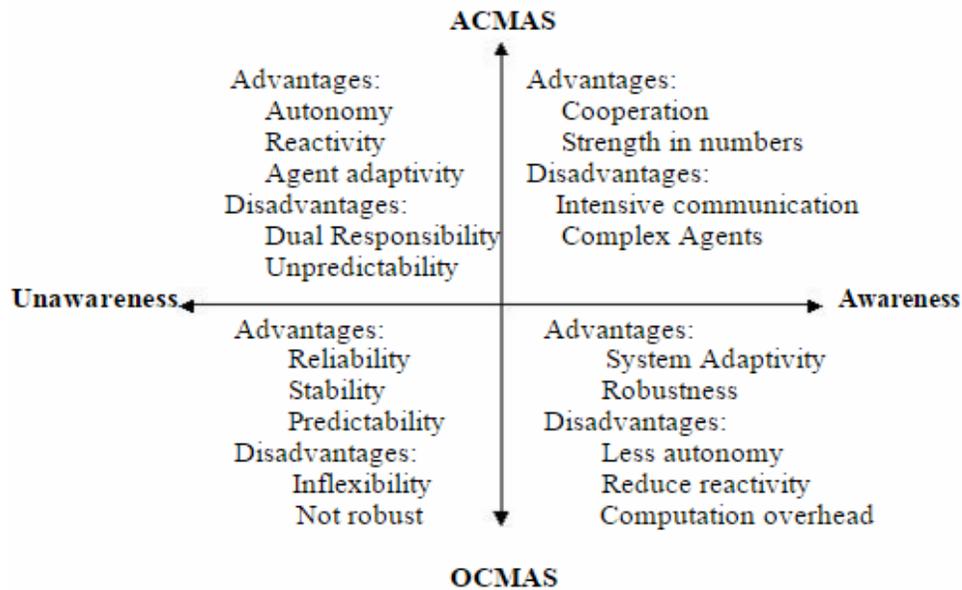

Figure 2: Advantages and Disadvantages of MOS-2 Quadrants

## 4. THE PROPOSED MOS-2 SPACE

Figure 3 presents the proposed MOS-2 space. As shown in the figure, the axes of a two-dimensional Cartesian system divide the space into four infinite regions, called quadrants, each bounded by two half-axes. In mathematics, these are often numbered from 1st to 4th and denoted by Roman numerals, lets take the same naming conversion, and thus the four quadrants can be identified as follows:

1. The I symbol identifies the ACMAS-Aware space quadrant
2. The II symbol identifies the ACMAS-Unaware space quadrant
3. The III symbol identifies the OCMAS-Unaware space quadrant
4. The IV symbol identifies the OCMAS-Aware space quadrant

Therefore, to position a MAS organizational model it should be studied and explored to see to which quadrant in the MOS-2 space it is best fit according to its characteristics and properties. If the considered model fits one of MOS-2 quadrants, then its position will be represented by a small circle as shown in Figure 4-a. But, if the model fits two quadrants, then its position is represented by a small oval shape expanded along the two space quadrants as demonstrated in Figure 4-b. Note that the oval part appeared in each quadrant should be relative to the extent to which the model realizes the characterizes of the MAS organization class represented by that quadrant. If the considered model realizes (partially or fully) the characteristics of three space quadrants then the position of this model can be represented as a half-circle expanded along the three space quadrants as shown in Figure 4-c. Finally, in case the considered model fits the whole space (it is rare but possible) then a circle expanded along the four space quadrants can be used to represent that perfect organizational model!, this case is demonstrated in Figure 4-d. Note also that the MOS-2 space can be enlarged to position simultaneously many organizational models. The next section applies the proposed space for positioning a number of familiar MAS organizational models.

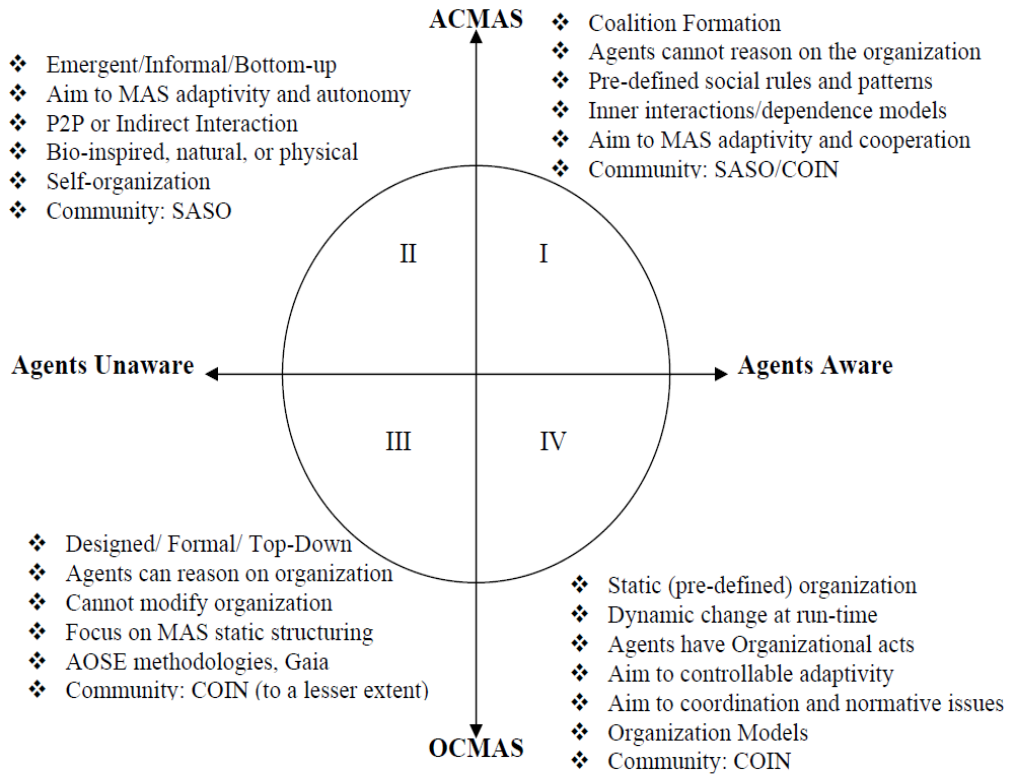

Figure 3: The MOS-2 space with the characteristics of each organization class

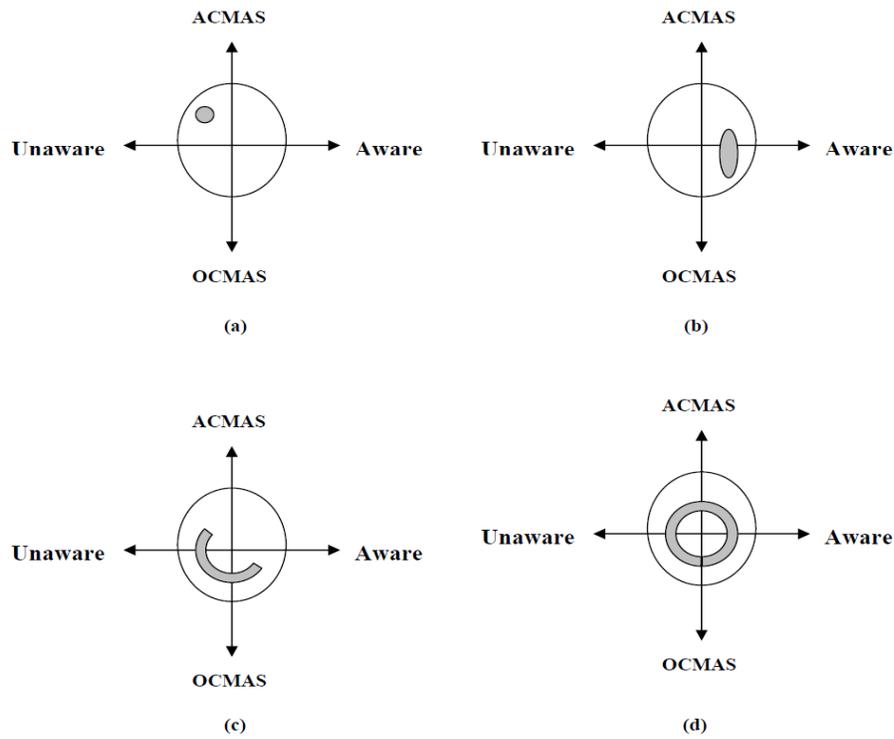

Figure 4: positioning of different types of organizational models

# 5. THE APPLICABILITY OF MOS-2 SPACE: CASE STUDIES

This section applies the MOS-2 space to position and compare a number of familiar MAS organizational models. The selected models are: AGR, MACODO, MOISE, Swarm-based approaches, Contract Net coordination model, and Gaia development methodology. Actually, there are a large number of MAS organizational models and approaches but we found that the selected ones are enough to demonstrate the usage of the proposed space and we leave to the reader the mission of trying to position any other model found in the literature or proposed by him.

## AGR

The AGR [8][9] is a MAS organizational model that adopts the OCMAS viewpoint. This model is influenced by both AOSE and social reasoning, in the sense that organization is used by designer to specify the system-to-be and by the agents that can dynamically perform organizational acts and possibly modify the organization. The designer uses abstract concepts such as Group Structure and Organizational Structure to specify application in design-time. The group structure is an abstract representation of the roles required in this group and their interaction relationships and protocols. The organization structure is the set of group structures expressing the design of a multi-agent organizational scheme. In run-time, the agents can reason on the organizational aspects and can modify the application structure by the dynamic creation of agents groups (agents partitioning) and dynamic forming of hierarchies of groups. Therefore, the AGR model fits well with the III and IV quadrants of MOS-2 space and can be positioned as shown in Figure 5-a.

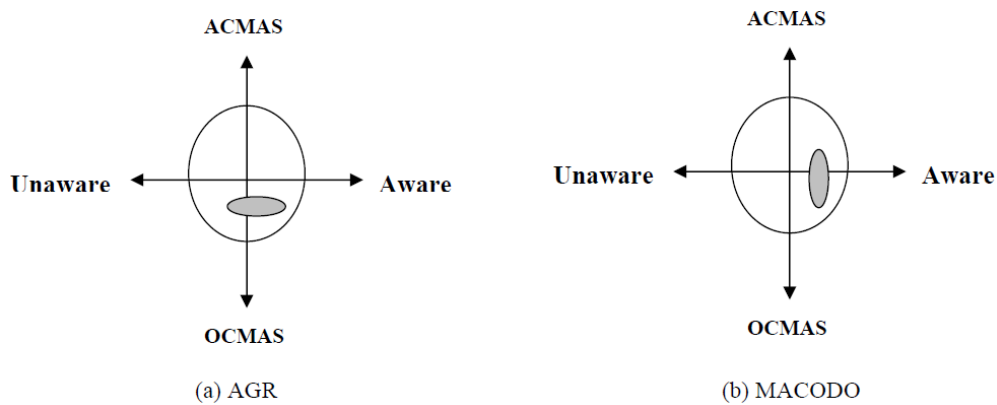

Figure 5: Positioning of AGR and MACODO organizational models

## MACODO

The MACODO [10][11] is a MAS organizational model that is really an interesting model because it tries to realize (although partially) both of the MAS engineering viewpoints: ACMAS and OCMAS. It, to a large extent, belongs to the OCMAS philosophy, which provides an explicit representation of agent organizations. Although it provides a formal predefined specification for system dynamic reorganization, it allows the agents to (according to the environment context) form a type of short-lived coalitions by presenting a cooperative behavior with each other. The agents' cooperative behavior is supervised and controlled by the organization controllers. The organizational model is part of an integrated approach, called MACODO (Middleware Architecture for COntext-driven Dynamic agent Organizations); in this model, the life-cycle management of dynamic organizations is separated from the agents, organizations are first-class citizens, and their dynamics are governed by laws. We see that, the MACODO organizational model is best fit with two space quadrants, IV and I of the MOS-2 space, as presented in Figure 5-b.

## MOISE

Hannoun et al. [12] proposed MOISE (Model of Organization for multI-agent SystEms); for modeling organizational aspects of MAS. The MOISE model is aimed at providing support in order to adapt an agent organization to its environment and to help it to efficiently achieve its goals. This model defines an organization which is composed by agents, roles, missions, and the deontic dimension. Each role represents a set of constraints that an agent follows when it plays this role. These constraints represent the structure dimension (relations between roles) and the functional dimension (missions, deontic dimension). A mission is a set of coherent goals that an agent can commit to. The deontic dimension specifies the permissions and obligations of a role in a mission. MOISE adopts a proactive reorganization carried out in a distributed way by monitor agents. The logic for reorganization is implemented at design time and cannot be changed during runtime. But the agents can modify the MAS organization according to the predefined logic. For example, the agents can change their roles or give a new obligation; a new role can be added to the system, etc. We see the MOISE model similar to the AGR model and can be positioned in the same way through quadrants III and IV of the MOS-2 space, as shown in Figure 6-a.

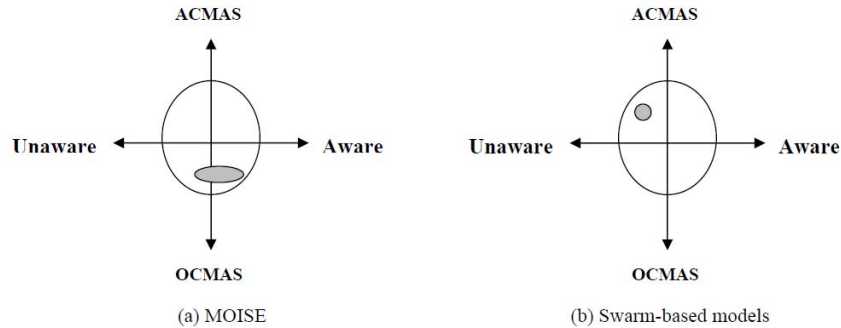

Figure 6: Positioning of MOISE and Swarm-based organizational models

### Swarm-Based Approaches

The swarm-based approaches adopt the ACMAS viewpoint where the agents are unaware of any higher level structure. The system organization is dynamic and informal, it is an emergent phenomena appears to the observer in the higher level as a result of the individual agents lower level interactions directly (in a peer to peer fashion) or indirectly (through environment). In this type of agent systems, the designer concerns only the individuals and the environments, and he doesn't give any attention to the global organizational level. When the designer develops an individual agent he put in his mind the application domain and its environment only. In these systems, the individuals are purely reactive that simply react to environment changes. The ant colony [13] represents a realistic natural example of these systems. In the ant colony there is no organizational behavior and constraints are explicitly and directly defined inside the ants. The main idea is that the organization is the result of the collective emergent behavior due to how agents act their individual behaviors and interact in a common shared and dynamic environment. This class of systems represents the pure ACMAS viewpoint, which is located in the II quadrant in the MOS-2 space, as shown in Figure 6-b.

### Contract Net

Contract Net Protocol (CNP) [14] is a task-sharing protocol in multi-agent systems, consisting of a collection of nodes or software agents that form a purposeful coalition. The agents are pre-augmented by the designer with some social rules, interaction models, and dependency models to be able to participate in purposeful coalitions. The organization is implicit and depends on the situation faced by the agents. The agents may be not able to reason about the global system organization, but they just follow the predefined social rules. The CNP model is best fit with the I-quadrant in the MOS-2 space, and can be represented by a small circle as shown in Figure 7-a.

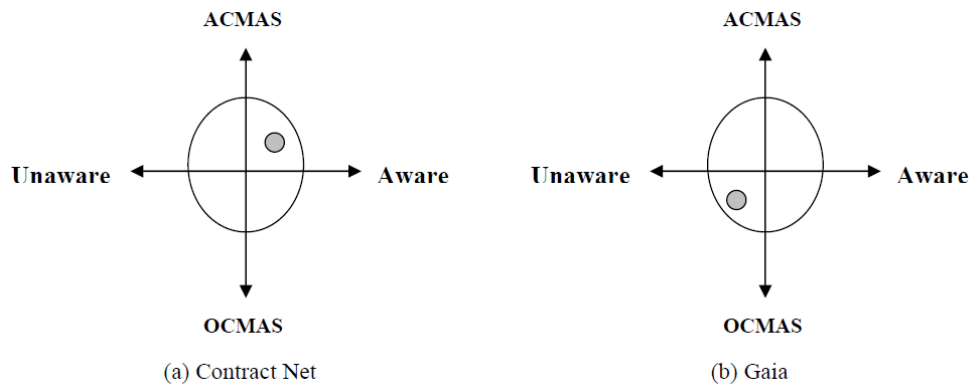

Figure 7: Positioning of Contract Net and Gaia organizational models

## GAIA

The Gaia [15] is an agent-oriented software engineering methodology (AOSE) [16]; it considers the system organization at the design-time. Organizations are specified before encoding the agents. Agents can reason on the organization at run-time but cannot be able of modifying it. In other words, In MAS that are modeled by the Gaia methodology, the system structure is defined in design-time and doesn't change in run-time (fixed structure). The Gaia-based MAS can be positioned in the III quadrant in the MOS-2 space, which represents the AOSE engineering approaches, as shown in Figure 7-b.

## NOSHAPE

The NOSHAPE MAS organizational model [17] is a recent model, although it is not matured yet, but it provides a novel approach for engineering complex and highly distributed MAS. Like the MACODO model, the NOSHAPE model tries to adopt the two MAS engineering viewpoints: the ACMAS and the OCMAS. The NOSHAPE model allows individual agents to loosely reshape the higher level system organization by emitting triggers (triggers can be seen as the pheromones released in the environment by ants in the ant colony as a type of indirect interaction). According to these triggers, the organizational level changes the system organization by establishing overlap relationships among higher order entities (agents' organizations, organizations' worlds, and worlds' universes). There are no any constraints imposed on the individual agents (except for mobility). The relationship between the individual agents' level and the organizational level is loose and depends only on the agents triggers. The organizational level can be seen as a helper or a guide to the agents. So, in the NOSHAPE model, the system structure emerges as a result of the agents triggers which are managed in a service-oriented manner, it is not possible to predict the next shape (structure) of the system, there are a pre-defined specification, and agents can modify indirectly the whole system structure by just emitting triggers. The NOSHAPE model aims to provide generality (relative to systems scale) and also aim to make the relation between the level of individual agents and the organizational level loosely coupled, so the agents can behave according to the ACMAS viewpoint (where agents are unaware of the organizational aspects) and the organizational level behaves according to the OCMAS viewpoint independently. Therefore, the NOSHAPE model can be positioned along the quadrants II, III, IV of MOS-2 space, as shown in Figure 8.

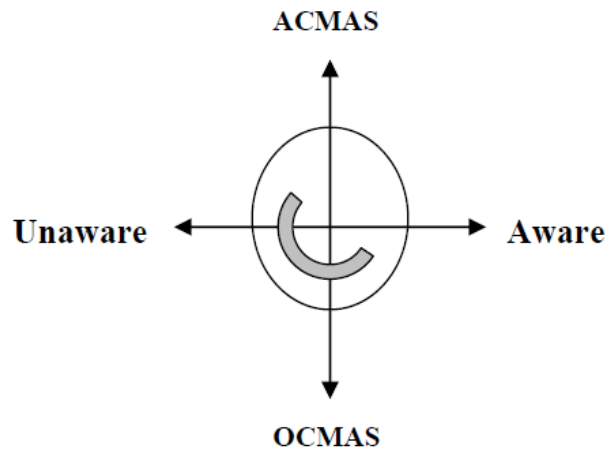

Figure 8: Positioning of the NOSHAPE MAS organizational model

## COMPLETE VIEW

All of the selected MAS organization models can be positioned and represented on one space diagram as shown in Figure 9. Based on the complete view shown in Figure 9, we can claim that the more space quadrants an organizational model visits, the more it has of features. In other words, it will possess the advantages of each space quadrants. Not only this, but also it will have the chance to match the disadvantages of one space quadrants to the advantages of another one. According to these results, we claim that the NOSHAPE model is a promising one because it visits three space quadrants.

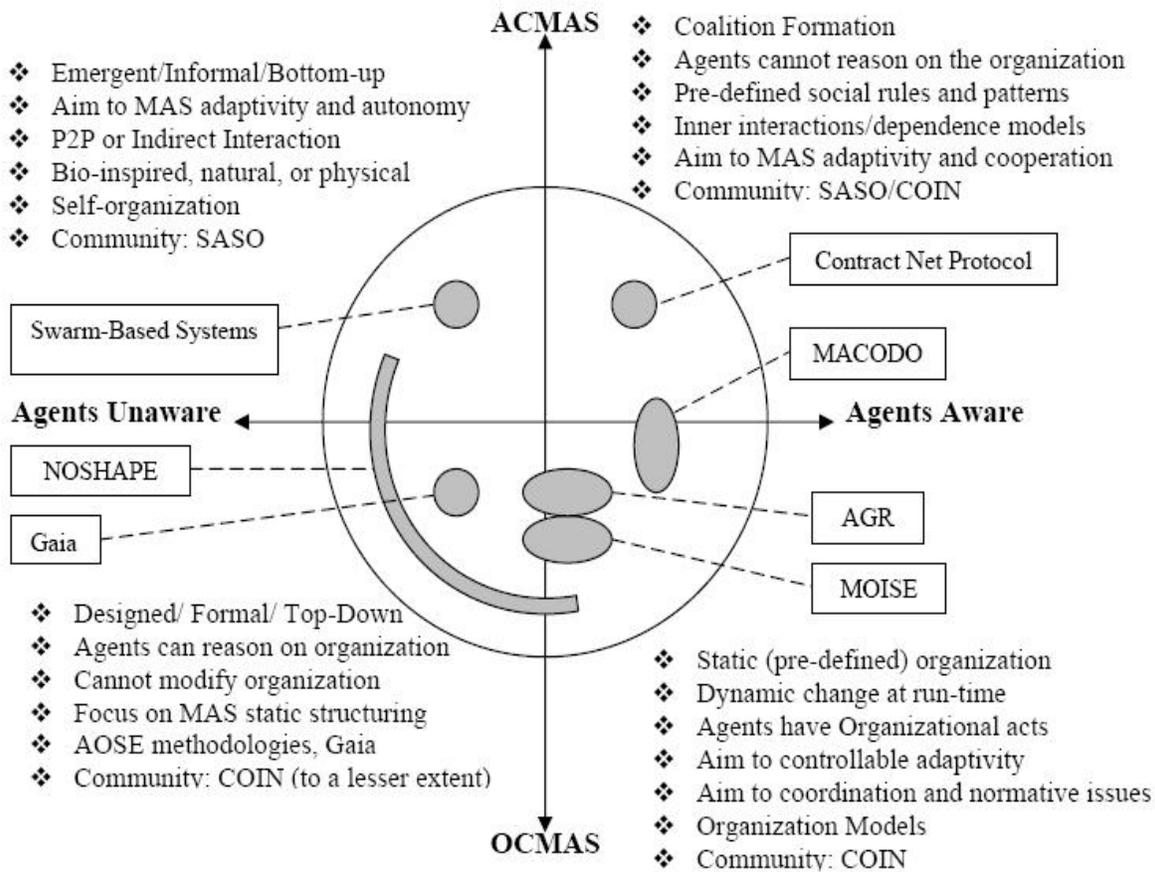

Figure 9: Complete view of positioning MAS organizational models

## 6. FUTURE TRENDS

The proposed MOS-2 space can be evolved to precisely position and compare MAS organizational models and approaches by considering the following issues:

1. Adding Time as a Dimension
2. Increasing the considered dimensions
3. The quantitative Approach

In the following subsections, these issues will be highlighted and some suggestions and ideas will be proposed to be addressed in future.

### Adding Time as a Dimension

Based on the proposed MOS-2 space for MAS organization classification, is it possible to ask this question "can an organizational model evolves dynamically and changes to a novel emergent model"? May be it seems like a pure fantastical idea but as we all know "science is not about why? It's about why not?" What this question means is to add a third dimension to the proposed space, it is Time, so it may seem logical to call it MOS-2T, let's illustrate this amazing idea! Consider the organization space shown in Figure 10-a, as shown there is a model $M_0$ positioned in the III quadrant (the AOSE quadrant), with the addition of time as a third dimension, lets denote the absolute position of a model as a tuple <P, T> where P is the observed position of the considered model on the space and T is the time where we observed the position, then:

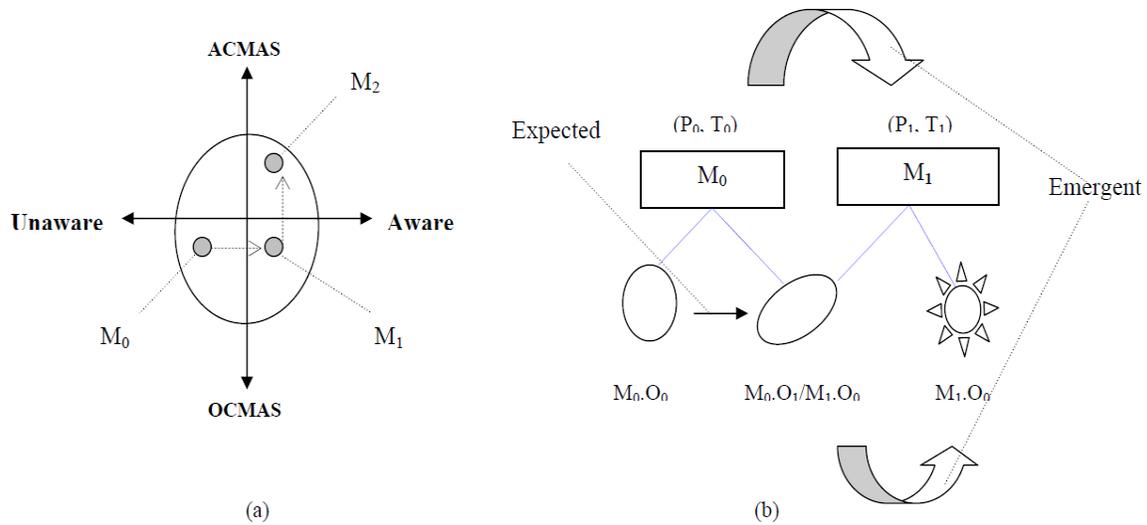

Figure 10: Dynamic evolution of organizational models

At $T_0$ the position of model $M_0$ is $<P_0, T_0>$, we may write it like:

$P(M_0) = <P_0, T_0>$  where $P: M \Rightarrow P \times T$  where $P = \{P_0, P_1, P_2, \ldots P_k\}$ or the set of possible positions. And $M = \{M_0, M_1, M_2, \ldots M_n\}$ is the set of possible models (assume it is a finite set).

If at time $T_1$ the model $M_0$ changed to a new emergent model $M_1$ (we can say that it is evolved, ignore questions like why? When? How? At least for now) then:

$P(M_1) = <P_1, T_1>$

And at time $T_2$ also $M_1$ evolved to $M_2$ as follows

$P(M_2) = <P_2, T_2>$   As demonstrated in Figure 10-a.

Also let's denote the organization of a MAS when the organization model $M_0$ is active at $T_0$ by

$O(M_0, T_0) = M_0.O_0$ where $O: M \times T \Rightarrow M.O$  where $M.O = \{M.O_0, M.O_1, \ldots, M.O_m\}$ or the set of possible organizations under the umbrella of a certain organizational model $M$.

So with $M_0$ is active, a dynamic organization change can happen as follows:

$M_0.O_0 \rightarrow M_0.O_1$

That is normal and expected according to the specification of the organization model $M_0$. But what is not normal is when the model $M_0$ is no longer active because it evolved to a novel emergent model:

$M_0 \rightarrow M_1$

As demonstrated in Figure 10-b, then it is expected to see a new organization of the system that is not planned by the designer and may violate the designer specifications, that can only happen if the agents are intelligent and have learning capabilities, so they may cause this amazing change in the organizational model as an emergent phenomena. As we see in Figure 10-b, after the dynamic emergent change of the organizational model takes place, the system organization $M_0.O_1$ relative to $M_0$ becomes $M_1.O_0$ relative the emergent model $M_1$, in other words $M_0.O_1$ becomes the initial organization of $M_1$, and the new model ($M_1$) will cause the system organization to change to a novel organization that is not planned by the designer. Dynamic evolution of pre-defined organizational models because of agents' intelligence and their ability to learn is expected to be the next mainstream research area in MAS discipline.

## Increasing the Considered Dimensions

The dimension of the proposed MOS-2 MAS organization space can be increased by adding other characteristics or properties of MAS organizational models. For example, a new dimension to represent the extent to which the model tackles the inter-reorganization and intra-reorganization can be added. The inter-reorganization is concerned with the organizational level interactions (i.e., interactions between groups of agents), and the intra-reorganization is concerned with individual agents interactions inside one group of agents. In this case the organization space will be a 3-dimension space as shown in Figure 11. Also in this case the name of space will be MOS-3 and if the time dimension is added, its name will be MOS-3T. The space shown in Figure 11 demonstrates how a model can be positioned in MOS-3 by considering three defined dimensions (x, y, z) correspond to classification criteria of MAS organizational models.

More dimensions can be added and the more dimensions added, the more fine-grained classification of MAS organizational models can be achieved. Therefore, the general name of the proposed space can be written as MOS-NT, where N is the number of used dimensions and T is the time dimension (the T should be removed if the time dimension is not considered).

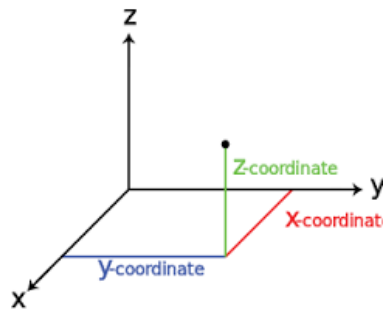

Figure 11: A 3-dimension space for MAS organization

## The Quantitative Approach

In the previous sections, we tackled MAS organizational models in a qualitative way. In fact, the qualitative positioning of MAS organizational models is largely limited by the imagination of the researcher. In the qualitative approach, the researcher depends on descriptions and observations, but sometimes it is better to provide a quantitative data based on rigorous measurements and calculations. So, it is possible to quantify the position of a MAS organizational model by finding numerically the extent to which the model realizes a certain dimension (i.e., ACMAS). Consider the MOS-2 space shown in Figure 12, what if we considered the central circle shown in the space as the unity circle, so the maximum value of a dimension is 1 and the minimum value is -1 (i.e., - ACMAS= OCMAS). Thus, when positioning a model, we should find a way to accurately calculate the extent to which the model can be considered to realize a certain dimension.

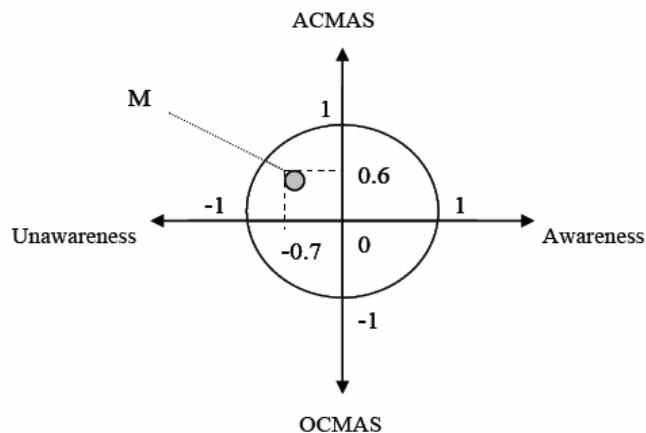

Figure 12: MOS-2 with a quantitative approach

Assume that there is a function V that is designed to calculate numerically the extent to which a model M is ACMAS/OCMAS. The function V can have the following signature:

V: M $\Rightarrow$ [-1,1]

Similarly assume that there is a function W that is designed to calculate numerically the extent to which the agents in a model M are Aware/Unaware. The function W can have the following signature:

W:M $\Rightarrow$ [-1,1]

Therefore the quantitative position of the model M in MOS-2 space can be determined as follows:

P(M)=(V(M), W(M))

The example in Figure 12 shows that:

V(M)= -0.7   and    W(M)= 0.6

So the position of the model M can be written as follows:

P(M) = (V(M), W(M)) = (-0.7, 0.6)

The problem now is how to implement the functions V and W. The complexity of these functions will increase as the number of considered dimensions increases. Not only this but also the number of this functions will increase because each considered dimension will need a function to quantify the extent to which this dimension is realized by the MAS organizational model. Moreover, their implementation will be more complex if the time dimension is considered. This research trend is highlighted as future work.

## 7. CONCLUSIONS

This paper proposes a new method for positioning and comparing of MAS organizational models based on a two dimension space, called MOS-2. MOS-2 uses two comparison aspects: the MAS engineering viewpoint (ACMAS/OCMAS) and the agents' awareness/unawareness of the existence of a higher organizational level. The proposed space provides a feasible, effective, visual, and semi-formal positioning method for comparing MAS organizational methods using an orthogonal coordinate system similar to the familiar Cartesian coordinate system. The applicability of the proposed MAS organization space have been demonstrated by using it to position and compare a number of familiar and recent MAS organizational models, other models can be positioned similarly. Future trends are highlighted and discussed in Section 7.